# Bifurcations Caused by the Diffusion-Induced Noise

Maria.K. Koleva* and L. A. Petrov

Institute of Catalysis, Bulgarian Academy of Science
1113, Sofia, Bulgaria
Fax: +359 2 9712967

*mkoleva@bas.bg

corresponding author: Maria K. Koleva

**Abstract**

The properties of the fluctuations large enough to induce bifurcations at open chemical systems at steady constraints are studied. The fluctuations that come from the diffusion-induced noise are considered. It is a generic for the surface reactions driving mechanism of fluctuations whose distinctive property is that it generates bounded fluctuations at any value of the control parameters. The range of boundedness is specific to a system but it is such that the system permanently stays within its thresholds of stability. This in turn ensures long-term stable evolution of the system. The frequency for occurrence of an induced bifurcation is carried out analytically. Its important property is that it gradually becomes insensitive to the reaction mechanism on increasing the excursion size.

**Introduction**

So far the interest to the open chemical systems exposed to a steady flow of reactants is focused mainly to the realisation of different dynamical regimes when the control parameters are varied. The bifurcation diagram of a system is determined by the reaction mechanism and the control parameters and it does not change in the course of time. The major assumption that provides such behavior is that the internal fluctuations are negligible around stable states. Their influence is enhanced to macroscopic effects only at bifurcation points (Nicolis and Prigogine, 1977). However, sharp sudden changes of different aspects of the reaction mechanism often occur in the course of time even at stable dynamical regimes (Scott, 1995). A possible explanation is that the changes are induced by large enough fluctuations when the catalyst properties remain unchanged. This is the case considered further. However, the established so far sources of internal fluctuations are such that the frequency for occurring of a fluctuation of necessary size is vanishingly small (Wentzel and Freidlin, 1984). The revealing of the contradiction is in looking for a new driving mechanism of fluctuations so that it can produce large enough fluctuations in a finite time interval even at stable dynamical regimes and far from bifurcations points. Recently one of us (Koleva, 1998) has introduced a new driving mechanism of internal fluctuations called diffusion-induced noise. It gives rise to macroscopic fluctuations at any value of the control parameters and even at asymptotically stable states. The diffusion-induced noise is common for all the surface reactions and is insensitive to the details of the reaction mechanism. Its generic property is that it gives rise to bounded fluctuations. The range of boundedness is specific to a system but it is such that the system permanently stays within its thresholds of stability. This in turn ensures long-term stable evolution of the system. On the contrary, i.e. if the fluctuations are unbounded, the system can blow up or get extinct in a finite time interval. Thus, the boundedness of fluctuations is a key to a stable long-term evolution of any natural and engineered system.

The boundedness renders any bounded irregular sequence (BIS) to be a subject of the Lindeberg theorem (Feller, 1970), i.e. it has finite mean and finite variance regardless to the details of its incremental statistics. Another obvious feature of BIS`es is that their increments are also bounded. Further, the case when the largest increment is much smaller than the threshold of stability is considered. Therefore, each large enough fluctuation can be approximated by an excursion: a trajectory of a walk originating at the mean value of a given sequence at time $t$ and returning to it for the first time at time $t + \Delta$.

In the next section it is found out that any deviation from the mean causes an effective change of the control parameters. Then, neither dynamical regime is anymore determined by merely the control parameters because the development of a large enough excursion at some instant can effectively "change" the control parameters so that to cause a bifurcation. Further considerations are based on the assumption that there is a diffeomorfism between the excursion size and the distance a bifurcation point. Thus, the bifurcation occurrence in the course of the time strongly depends on the properties of the excursions such as the size, duration, frequency of occurrence. The questions considered next are: what is the relation between the size $A$ and the duration $\Delta$ of an excursion. What is the frequency of occurrence of an excursion of a given size. Are successive excursions well separated and if so, what prevents their overlapping.

It turns out that the boundedness provides that each excursion is loaded in an "embedding" time interval whose duration is exerted as a multi-valued function in the course of the time. The duration of any embedding time interval $T$ is always greater than the duration of the excursion itself $\Delta$. This provides the separation of the successive excursions by quiescent time intervals. It is worth noting, that the presence of embedding time intervals preserves the boundedness since it does not allow an arbitrary grow of the excursion size on overlapping the successive excursions. The relation between the range of the duration of the embedding time intervals $T$ and the duration of the corresponding excursion $\Delta$ is carried out explicitly.

The further considerations are made under the approximation of the large fluctuations by excursions created by the fractal Brownian motion of the increment walk. The latter renders the following relation between the size of the excursion $A$ and its duration $\Delta$: $\sqrt{\langle A^2 \rangle} \propto \Delta^{\beta(A)}$, $\beta(A)$ is set on the particularities of the incremental statistics; the averaging is over the samples realisations. This gives rise to the question about the dependence of the relations $T \leftrightarrow \Delta \leftrightarrow A$ on the particularities of the incremental statistics. It should be stressed that the details of the latter are explicitly related to the details of the reaction mechanism (sec.1). Now we come to the major question of the present paper: what is the dependence of the excursions properties on the incremental statistics (i.e. on the reaction mechanism). It is found out explicitly that on increasing of the excursion size there is a gradual decrease the influence of the incremental statistics on $T, \Delta$ and $A$.

The paper is organised as follows: the driving mechanism of the diffusion-induced noise and its mathematical description is presented in the next section. It is pointed out the association between the incremental statistics and the reaction mechanism. The explicit form of the relations $T \leftrightarrow \Delta \leftrightarrow A$ and their dependence on the incremental statistics is studied in sec.2. Experimental evidences that visualise the effective change of the control parameters under fluctuations are presented in sec.3.

# 1. Diffusion-Induced Noise

The theory of the diffusion-induced noise is built upon the idea about a diffusion-induced non-perturbative interaction. The notion of the latter is defined when a Hamiltonian changes qualitatively under certain perturbation(s). The changes involve opening or closing of output channel(s). These interactions are not exotic occasions but they are even necessary for proceeding of any reaction. To elucidate the statement one should refer to the notion of stoichometry. When concerning a reaction elementary step it implies that the latter is initiated if and only if certain species of required types and states form certain local configuration. Then, the Hamiltonian of each of the reactants becomes unstable to the perturbations induced by the other species. The process leads eventually to the formation of the reaction products that are different from the reactants. Therefore, the *ad hoc* assumption implies that the initial Hamiltonians changes qualitatively during reaction (apart from that of a catalyst species). The necessity of a certain local configuration for driving any reaction event is immediately related to the boundedness of matter and energy involved locally. When the Hamiltonian of an interaction changes qualitatively after involving an alien to a given local configuration species and when it happens due to the diffusion of that species, the non-perturbative interaction is called a diffusion-induced one.

Next we consider the diffusion-induced non-perturbative interactions at surface reactions when a surface is exposed to a steady flow of reactants. It will be shown next that diffusion-induced non-perturbative interactions appear in the process of chemisorption. The latter is a process, typical for all surface reactions since a reaction event is available only between already chemisorbed species of required types. Therefore, the noise behavior is insensitive to the details of the reaction mechanism.

The gas/surface system is approximated by the ideal adsorption layer approach introduced by Langmuir. Besides its apparent simplicity it allows the system to be considered as a spatially homogeneous one.

The physical background of the notion of the active site is that a specific for each gas/surface system potential well is formed at certain places on the surface, called active sites. When a gas species hits the surface at an empty active site, it may lose enough kinetic energy to be trapped in one of the highly excited bound states of the well. The further relaxation to the ground (chemisorbed) state involves several steps ascertained by the energy loss mechanism specific to the system. The form of the potential well and the energy loss mechanism determines the characteristics of any relaxation. There is, however, an important general property of any relaxation that is: at any excited level there are four open channels. An excited species can: (i) relax to a lower level, it can be scattered inelastically to a higher level, it stays at the same level at the same site or it migrate to a next site. The probabilities for any of these events at any level can be evaluated by a quantum-mechanical approach appropriate for the system. Further in the paper this relaxation is called ordinary relaxation. It, however, can be interrupted in any instant and at any level by an adspecies that reaches that active site by migration. Since no more than one species can be chemisorbed on a single active site, the *ad hoc* assumption states that the involving of another adspecies in the process of relaxation changes the Hamiltonian qualitatively - from attractive it becomes repulsive. Here we encounter the physical meaning of the boundedness of fluctuations: the microscopic stability of the system considered is presented by the fact that no more than one species can be chemisorbed at a single active site. The qualitative change of the Hamiltonian ensures that property and it is achieved by involving a *finite* energy (it does not exceed the adsorption energy) and *finite* matter ( a single species). Thus, the involved matter and/or energy is always bounded. Therefore, the excited species can only be scattered to a higher level or migrate to another active site. It can complete the relaxation only if, after migration, it reaches an empty active site. The adsorption rate, however, changes non-smoothly compared to the rate of the ordinary relaxation. The ordinary relaxation can be interrupted by another adspecies at any level, since there is no special moment for that adspecies to occur at the active site. In other words, since the relaxation and the diffusion are processes considered independent of one another, a non-perturbative interaction can happen at any level of ordinary relaxation with equal probability. Thus, we come to an example of what we called above a diffusion-induced non-perturbative interaction. Therefore, the adsorption rate becomes a multi-valued function, each selection of which corresponds to one of the levels at which a non-perturbative interaction is carried out. The values of the different selections are specific for the system and they can be evaluated by an appropriate quantum-mechanical approach. Under the assumption about the local onset of fluctuations only one selection, randomly chosen among all possible, appear at a given instant at a given active site. It is obvious that at next instant another selection can appear with the same probability. Thus, the temporal variation of any local adsorption rate forms a bounded irregular sequence in the time course. It should be stressed that the probability for undergoing a single diffusion-induced non-perturbative interaction is proportional to the concentration of the adspecies. So, the probabilities for an ordinary relaxation and for a relaxation through undergoing a single diffusion-induced non-perturbative interaction are of the same order.

The above driving mechanism of fluctuations can easily be adapted to any surface reaction because its major features, namely, surface migration of the reactants and the fact that no more than only certain number of species specific to the system can be in ground state locally, are common properties of the reaction systems robust to their nature.

It has been proven (Koleva, 2002) that the global adsorption rate always equals certain local adsorption rate. The proof is not trivial because the lack of correlation among local diffusion-induced non-perturbative interactions (respectively lack of correlation among local adsorption rates) induces a spatial non-homogeneity even in the academic case of identical mobility and adsorption properties of all the reactants and intermediates. The induced spatial non-homogeneity is permanently sustained by the lack of coherence between the trapping instants and the adspecies mobility. In turn, the adlayer configuration would vary in uncontrolled way which in short time would cause either the reaction termination or the system breakdown. The elimination of the spontaneous spatial non-homogeneity is achieved at the expense of new assumptions about the Hamiltonian response to the disturbances introduced by the uncorrelated mobility of the adspecies. One of these assumptions is that the Hamiltonian for adsorption (reaction) separates in a rigid and flexible part. For the sake of simplicity only adsorption is considered next. At the rigid part the impact of the adspecies can be treated as perturbations while at the flexible part their impact is compatible to the "forces" that creates the Hamiltonian. Then, the spectrum of the rigid part has a specific to the system discrete spectrum. It is proven that flexible part has certain chaotic properties and thus it is insensitive to the details of the rigid Hamiltonian. Berry (Berry, 1985) has proven that the spectrum of any chaotic Hamiltonian obeys Wigner distribution. Thus the spectrum of any Hamiltonian comprises two parts: the low-lying one that comes from the rigid part and is specific to the system and the upper part that is universal in the sense that it is insensitive to the particularities of the rigid part of the Hamiltonian (respectively to the system). Another assumption put forth is that the interactions among trapped in the chaotic states species relax through excitations of the local acoustic phonons. In turn, the latter contribute to the disturbances that create the chaotic part of the Hamiltonian. This allows a creation of a non-local feedback that acts towards "synchronising" of the local adsorption rates. After the synchronisation is completed, the relaxation proceeds through the rigid part of the spectrum. After the chemisorption is completed, starts the reaction and then again the adsorption: first synchronization and then relaxation trough rigid part of the spectrum and so on. The global adsorption rate has two distinctive properties: (I) its value varies with different adsorption cycles; (ii) at each cycle it equals that local adsorption rate which is in the most favorable local energetic environment. It turns out that at successive adsorption cycles different selections are in the most favorable local energetic environment. Since the most prominent variations of the adsorption rate are those that involve a diffusion-induced non-perturbative interactions, the current value of the global adsorption rate is randomly chosen among all the available selections of the individual adsorption rate.

Thus, the inevitable presence of the diffusion-induced non-perturbative interactions makes the global rate a multi-valued function at each value of the control parameters. The stochastic behavior appears from the permanent random choice among selections because only one selection takes place at every instant. Therefore the macroscopic kinetic equations read:

$$\frac{d\vec{X}}{dt} = \vec{\alpha}\hat{A}_{det}(\vec{X}) - \vec{\beta}\hat{R}_{det}(\vec{X}) + \vec{\alpha}\hat{\mu}_{ai}(\vec{X}) - \vec{\beta}\hat{\mu}_{ri}(\vec{X}), \qquad (1)$$

$\vec{\alpha}$ and $\vec{\beta}$ are these parts of the rates that explicitly depend on the external constraints and thus, for steady conditions, appear as control parameters in eqs.(1) ; $\vec{X}$ is the vector of the reaction species concentrations; $A_{det}(\vec{X})$ ($R_{det}(\vec{X})$) are sums of the rates for the ordinary adsorption (ordinary reaction) and the average values of all the possible selections at the adsorption (reaction). The subscript $i$ serves to stress that only one selection, randomly chosen among all the possible, takes place at a given instant. The definition of $\mu_{ai}(\vec{X})$ and $\mu_{ri}(\vec{X})$ determines their general properties that are very important and play a crucial role. At first, their definition resolves strong dependence of their values and statistics on the reaction mechanism. Another important property is that $\mu_{ai}(\vec{X})$ and $\mu_{ri}(\vec{X})$ are zero-mean and bounded in the range [-1,1]. Other crucial feature is that the selection appearance has Markovian property in the sense that the probability for occurrence of a given selection does not depend on the probability for the appearance of any selection at the previous instant. The boundedness and the Markovian property of the selection appearance immediately yields the separation of the solution of eqs.(1) to a deterministic and a stochastic part (Koleva and Covachev, 2001). The deterministic part is set through the solution of:

$$\frac{d\vec{X}_{det}}{dt} = \vec{\alpha}\hat{A}_{det}(\vec{X}_{det}) - \vec{\beta}\hat{R}_{det}(\vec{X}_{det}) \qquad (2)$$

Eqs.(2) is a system of ordinary differential equations and thus it defines the bifurcation diagram. However, the interplay between the determinism of eqs. (2) and the fluctuations in eqs.(1) causes permanent deviations from any dynamical regime prescribed by eqs.(2). Formally, any difference $(\vec{X}(t) - \vec{X}_{det})$ can effectively be presented as a result of a change of the control parameters. The latter immediately causes change either of the characteristics of the dynamical regime or it even causes change of the type of the dynamical regime. It is obvious that any induced bifurcation needs a fluctuation of certain size. Thus, the problem about the frequency for occurrence of a fluctuation of a given size becomes crucial for the induced bifurcations. This is the task of the next section.

It is obvious that the stochastic part of the solution of eqs.(1) $(\vec{X}(t) - \vec{X}_{det})$ is always a zero-mean bounded irregular sequence. The increments between its successive terms are set by $\mu_{ai}(\vec{X})$ and $\mu_{ri}(\vec{X})$. This explicitly makes the incremental statistics to be the bearer of the particularities of the reaction mechanism. A property insensitive to the incremental statistics is that all the increments are bounded. The range of boundedness is specific to the reaction mechanism and $X_{det}$ but it is permanently finite. Actually this property justifies the approximation of the large fluctuations by excursions. The meaning of the boundedness of the increments is that any local change of the energy and/or matter should not exceed certain threshold(s) such that the system stays permanently stable.

## 2. Statistics of the Excursions

The task of the present section is to evaluate explicitly the characteristics of the excursions: their amplitude, duration and frequency for occurrence. The importance of this question is rendered by the association of the distance to a bifurcation point with the excursion size. The study of the induced bifurcations becomes very important in the sense that the fluctuations cause by the diffusion-induced noise are inevitable: its presence is insensitive both to the catalyst preparation and the reaction mechanism and to the values of the control parameters (e.g. partial pressures of the reactants and/or the temperature). On the other hand, the fluctuations can cause not only bifrcations, but as well they can change the reaction route, can trigger phase transitions etc.

Our first task is to work out explicitly the relations $T \leftrightarrow \Delta \leftrightarrow A$, i.e. the relations between the duration of the embedding time interval $T$, respectively the duration $\Delta$ and the amplitude of the corresponding excursion $A$.

All further considerations are based on the assumption that the incremental correlations have finite radius smaller than the threshold of stability of the system. This assumption provides an arbitrary long-term stable evolution of the system considered. Indeed, if the radius of incremental correlations becomes larger than the threshold of stability certain fluctuations can cause the system breakdown.

### 2.1. Relation $T \leftrightarrow \Delta$

Further it is proven that the relation $T \leftrightarrow \Delta$ is imposed by the boundedness and is insensitive to the increment statistics. The presence of the "embedding" intervals renders that any successive excursions are separated by non-zero quiescent intervals. This does not allow the overlapping of the successive excursions and thus prevents the grow of the excursion amplitude to an arbitrary size. Thus, the "embedding" preserves the boundedness arbitrarily long-term. The present task is to establish the relation between the duration of the embedding intervals and the duration of the corresponding excursions. That relation is based on the notion of an excursion: a trajectory of a walk originating at the mean value of a given sequence at time $t$ and returning to it for the first time at time $t + \Delta$. Therefore, the probability for an excursion of duration $\Delta$ is determined by the degree of correlation between any two points of a sequence. On the other hand, the probability that any two points of a sequence, separated by distance $\eta$, have the same value is given by the autocorrelation function $G(\eta)$. A generic property of the BIS`es (Koleva and Covachev, 2001) is that the autocorrelation function of any of them can be defined for sequences of arbitrary but finite length $T$. Yet, its shape is universal, namely:

$$G(\eta, T) \propto 1 - \left(\frac{\eta}{T}\right)^{\nu(\eta/T)}. \tag{3}$$

where $\nu\left(\dfrac{\eta}{T}\right)$ is an continuous everywhere monotonically decreasing between the following limits function:

$$\nu\left(\dfrac{\eta}{T}\right) \to p-1 \quad \text{as} \quad \dfrac{\eta}{T} \to 0 \tag{4a}$$

$$\nu\left(\dfrac{\eta}{T}\right) \to 0 \quad \text{as} \quad \dfrac{\eta}{T} \to 1. \tag{4}$$

Then, the probability that an excursion of duration $\Delta$ happens in an interval $T$ reads:

$$J = \int_a^b x^{\pm\alpha(x)} dx = \dfrac{b^{\alpha(b)+1}}{1 \pm \alpha(b)} - \dfrac{a^{\alpha(a)+1}}{1 \pm \alpha(a)} . \tag{6}$$

Thus,

$$P(\Delta, T) = \dfrac{1}{T}\int_0^\Delta G(\eta, T)d\eta = P(\Delta, T) = \int_0^{\Delta/T} \left(1 - \varepsilon^{\nu(\varepsilon)}\right)d\varepsilon \tag{5}$$

By the use of the standard calculus (Koleva, 2002), the integration of any power function of non-constant exponent $\alpha(x)$ ($\alpha(x)$ is an everywhere continuous function) reads

$$P(\Delta, T) = \dfrac{\Delta}{T}\left(1 - \left(\dfrac{\Delta}{T}\right)^{\nu\left(\frac{\Delta}{T}\right)}\right). \tag{7}$$

The $P(\Delta, T)$ dependence only on the ratio $\Delta/T$ verifies the assumption that every excursion of duration $\Delta$ is "embedded" in an interval of duration $T$ so that no other excursion happens in that interval.

The next step is to work out the shape of $P(\Delta, T)$. Its role is crucial for the behavior of the excursion sequences. To elucidate this point let us consider the two extreme cases:

- $P(\Delta, T)$ is a sharp single-peaked function. It ensures a single value of the most probable ratio $\dfrac{\Delta}{T}$. In other words, the duration of the most probable embedding interval associated with an excursion of the most probable duration $\Delta_0$ has single value $T_0$ such that $\dfrac{\Delta_0}{T_0}$ is the peak value of $P(\Delta, T)$. Then, most probably the excursion sequence matches a periodic like behavior.

- $P(\Delta, T)$ has a gently sloping maximum. Then, the relation between $\Delta$ and $T$ behaves as a multi-value function: a range of nearly equiprobable values of $T$ corresponds to each most probable $\Delta$. In the course of the time this produces a random choice of the duration of the "embedding" intervals even when the sequence comprises nearly identical excursions. Then, the excursion sequence has a non-periodic behavior.

The establishing of the particular shape of $P(\Delta, T)$ requires the knowledge about the explicit shape of $\nu\left(\dfrac{\Delta}{T}\right)$. Next it is worked out on the grounds of the proof that neither BIS sustained to an arbitrary length comprises any long-ranged increment correlations. The general restriction on the increment correlation size requires an uniform contribution to the power spectrum of all scales, i.e. there are no "special" frequencies at the power spectrum.

The only factor that can modify the power spectrum is the non-constant exponent $\alpha(f)$ of its shape $1/f^{\alpha(f)}$. The boundedness requires the monotonic decay of $\alpha(f)$ in the limits $[1,p]$ not specifying its shape (Koleva and Covachev, 2001). Our task now is to establish the shape(s) of $\alpha(f)$ that fits the lack of long-range increment correlations. In virtue of the strict monotony of the power spectrum the required criterion is that neither any its component nor any its derivative of arbitrary order has a specific contribution. Simple calculations yield that it is achieved if and only if the shape of $\alpha(f)$ is the linear decay, namely:

$$\alpha(f) = (1 + \gamma f) \tag{8}$$

Eq. (8) provides that the $1/f^{\alpha(f)}$ derivative of an arbitrary order has the same sign throughout the entire frequency interval of the power spectrum. On the contrary, for any non-linear decay of $\alpha(f)$ each order derivative of $1/f^{\alpha(f)}$ changes its sign at certain frequencies. Thus, only the linear decay does not introduce any additional scale to those inherent for the increment statistics.

Because of the diffeomorfism between $\alpha(f)$ and $\nu\left(\dfrac{\Delta}{T}\right)$ it is obvious that the shape of $\nu\left(\dfrac{\Delta}{T}\right)$ reads:

$$\nu\left(\frac{\Delta}{T}\right) = (p-1)\left(1 - \frac{\Delta}{T}\right). \tag{9}$$

The plot of $P(\Delta, T)$ with the above shape of $\nu\left(\dfrac{\Delta}{T}\right)$ shows that it has a gently sloping maximum: indeed, the values of $P(\Delta, T)$ in the range $\dfrac{\Delta}{T} \in [0.25, 0.4]$ vary by less than 7%. Outside this range $P(\Delta, T)$ decays sharply. Thus, though $P(\Delta, T)$ is a single-valued function, it provides a multi-valued relation between the most probable values of $\Delta$ and $T$, namely: a certain range of nearly equiprobable values of $T$ is associated with each $\Delta$. In the course of the time the multi-valued relation is exerted as a random choice of the duration of the "embedding" intervals even when the sequence comprises identical excursions.

## 2.2. Relation $A \leftrightarrow \Delta$. Symmetric Random Walk as the Global Attractor for the Fractal Brownian Motion

Every bounded irregular sequence (BIS) comprises fine-structure fluctuations superimposed on the large-scaled ones. The former are created by the short-range increment walk and the latter appear as successive excursions from the mean value of the BIS. It is to be expected that the fine structure fluctuations strongly depend on the particularities of the increment statistics while the large scaled ones have certain generic properties robust to that statistics. One such property is that there is certain uniform relation between the amplitude of each fluctuation and its duration. Next it is worked out for the case when the increment size is much smaller than the thresholds of stability. Then the fluctuations can be approximated by excursions created by continuous fractal Brownian motion of the increment walk in the course of the time. The latter renders certain relation between the amplitude $A$ and the duration $\Delta$ of the excursion, namely: $\sqrt{\langle A^2 \rangle} \propto \Delta^{\beta(\Delta)}$, $\beta$ is set by the particularity of the increment statistics; the averaging is over the sample realisations. The dependence of $\beta$ on $\Delta$ comes from the interplay between the finite radius of the increment correlations $a$ and the amplitude of the excursion itself that is limited only by the thresholds of stability.

Next it is proven that whenever $a << A$ the fractal Brownian motion uniformly approaches the symmetric random walk with a constant step. Indeed, since the increment size is permanently bounded, the mean square deviations (m.s.d.) of all the trajectories of the same number of steps $N$ is confined in a finite range. Therefore, these m.s.d. also form a BIS. Further, according to the Lindeberg theorem (Feller,1970) the latter has finite mean and finite variance and thus it is a subject to the Central Limit Theorem. As a result, the m.s.d. are Gaussianly distributed. In turn, the sizes of the successive trajectories of a given $N$ are predominantly equal to the corresponding mean. So, the increment correlations are to be associated with a single trajectory whose number of steps $m$ is related to the increment correlation size and the exponent $\tilde{\beta}$ is specific to the increment statistics. In other words, the increment correlations create "blobs" whose size is $\sqrt{\langle a^2 \rangle} \propto m^{\tilde{\beta}}$. Then, the large excursions can be approximated by a symmetric random walk with constant step equal to the size of the blobs. Thus the dependence of any large scale excursions on its duration reads:

$$\sqrt{\langle A^2 \rangle} \propto N^{0.5} m^{\tilde{\beta}} \qquad (10)$$

where $N$ is the number of the blobs.
It is obvious that when $N >> m$ the dependence tends to:

$$\sqrt{\langle A^2 \rangle} \propto N^{0.5} a \qquad (11)$$

where $a$ is considered constant independent of $N$. So, the symmetric random walk with constant step appears as the global attractor for any fractal Brownian motion.

It is worth noting that the finite size of the "blobs" ensures the uniform convergence of the average to the mean of the original BIS. Indeed, the distinctive property of any fractal Brownian motion is that any power $\beta \neq 0.5$ arises from an arbitrary correlation between the current increment $\mu_i$ and the corresponding step $\tau_i$. Then the average $A$ reads:

$$A = \sum_{i=1}^{N} \mu_i(\tau_i)\tau_i = \sum_{i=1}^{N} (-1)^{\gamma_i i} \tau_i^{\beta_i} \qquad (12)$$

and correspondingly the m.s.d. :

$$\langle A^2 \rangle \propto \left\langle \sum_{i=1}^{N} (\mu_i(\tau_i)\tau_i)^2 \right\rangle = \left\langle \sum_{i=1}^{N} \tau_i^{2\beta_i} \right\rangle \qquad (13)$$

where the averaging is over the different samples of the trajectory; The property of the above relations is that whenever the probabilities for $\gamma_i$ odd and even are not permanently equal there is a correlation between the increment and the corresponding step. So $A$ is certainly non-zero that immediately makes that the deviation from the mean non-zero. Moreover, eq.(12) yields that $A$ can become arbitrarily large on increasing $N$. On the contrary, a permanent equal probability for $\gamma_i$ odd and even means independence from one another of the increments and the steps. It yields $A = 0$ which guarantees the uniform convergence of the average to the mean.

### 2.3 Frequency of Occurrence

Our first task is to elucidate that the excursion appearance meets the three generic features of a Poissonian process, namely: (i) the excursion appearance is a stationary process. (ii) the successive excursions are independent from one another events; (iii) no more than one event can be developed at any instant. The stationarity of the excursion sequences is guaranteed by the following interplay between the boundedness and the lack of any long-ranged increment statistics. Indeed, according to the Lindeberg theorem (Feller, 1970), any BIS has finite mean and finite variance insensitively to the increment statistics. Further, the lack of long-ranged increment correlations guarantees the uniform convergence of the average to the mean. This, in turn verifies the stationarity of the excursion appearance. This along with the embedding of each excursion in a larger interval is the warrant that the successive excursions are independent from one another events.

On the other hand, both the boundedness and the lack of long-ranged increment correlations render that every BIS is a subject to the central limit theorem. So, the departure from the mean obeys the Gaussian distribution. Our task now is to elucidate the interplay between the Poissonian properties and the Gaussian ones. When we are interesting in the frequency for an excursion of a given size $A_0$, all the excursions of the amplitude less then the size $A_0$ effectively contribute to the quiescent intervals. Then, the excursion of a given amplitude appears in the time course with the probability that reads:

$$P(A) = \frac{A^{1/\beta(A)} \exp(-A^2/A_\sigma^2)}{(1+\eta)A^{1/\beta(A)} \exp(-A^2/A_\sigma^2) + \int_0^{A_{0c}} A^{1/\beta(A)} \exp(-A^2/A_\sigma^2) dA} \qquad (14)$$

The required probability $P(A)$ is given by the ratio between the weighted duration $A^{1/\beta(A)} \exp(-A^2/A_\sigma^2)$ of an excursion of amplitude $A$ and the weighted length of the corresponding embedding interval. The latter comprises the weighted duration of the excursion itself, the weighted duration of the natural embedding interval expressed through $\eta$ and the weighted duration of the quiescent interval due to coarse-graining $\int_0^{A_{c0}} A^{1/\beta(A)} \exp(-A^2/A_\sigma^2) dA$; $A_\sigma$ is the variance of the BIS and in the present consideration it is a parameter. The Poissonian properties of the excursion appearance ensure that $P(A)$ has the same value at every point of the sequence. It should be stressed that eq.(14) holds only for $A \geq A_0$. Otherwise $P(A) \equiv 0$ since all the excursions of amplitude less than the size $A_0$ effectively contribute to the quiescent intervals.

The behavior of any BIS is inherently related to the incremental statistics trough the explicit dependence of $P(A)$ on $\beta(A)$ in eq.(14). However, when the size of the required excursion is much larger than the size of the symmetric random walk, $\beta(A)$ turns constant equal to $0.5$. Then, $P(A)$ gradually gets insensitive to the details of the increment statistics. Thus, whenever $a << A_0$ the behavior of the excursions of the size $A >> a$ becomes universal. Since $a$ is finite, there is always large enough size such that $a << A_0$. So, at large enough amplitudes the excursion behavior always becomes insensitive to the reaction mechanism.

The case of primary importance is $a << A_0 << A_\sigma$ because the universality comes along with large value of the frequency for excursion occurrence. Indeed, in this case eq.(14) becomes:

$$P(A_0) \approx \frac{1}{1 + \frac{A_0}{A_\sigma}} \propto 1 \qquad (15)$$

## 3. Comparison to the Experiment

The goal of the section is a perceptible presentation of the drastic effective change of the control parameters under large excursions. 3 successive parts of equal length of a time series recorded at oxidation of HCOOH over supported Pd catalyst are presented at Fig. 1(a), 2(a) and 3(a). The corresponding power spectra are presented at Fig. 1(b), 2(b) and 3(b). It is well known that the particularities of any irregular time series are best revealed in the corresponding power spectra.

The purpose for presenting namely that time series is twofold. On the one hand, there is an excursion of a considerable size: Fig. 1(a). On the other hand, the dynamical regime is a limit cycle that is characterised by the discrete band that is presented at all 3 power spectra. The discrete band exhibits the following properties: the period of the limit cycle is the same at all 3 power spectra while its amplitude reveals a large sensitivity to the fluctuations size (it is different at all 3 power power spectra). The most prominent difference is manifested at the presence of a large excursion (Fig.1(a)): it is visible that the limit cycle amplitude at Fig1(b) is about 10 times smaller than that of the limit cycles at Fig. 2(b) and 3(b). It is also visible that the amplitude of the limit cycle is slightly different in Fig. 2(b) and 3(b) though no considerable differences are visible at Fig.2(a) and Fig.3(a). It is well known that the amplitude of a limit cycle is very sensitive to the values of the control parameters on the contrary to the period. This is exactly what we have: a large sensitivity of the limit cycle amplitude to the effective change of the control parameters and a robustness of the limit cycle period in the same time. Therefore these results support our suggestion that the fluctuations do cause effective deviations of the control parameters in the course of time.

Another purpose for presenting that time series is to ensure that the large fluctuation at Fig. 1(a) is caused by the diffusion-induced noise. A strong evidence in favour are the features of the continuous band at all 3 power spectra. Previously we (Koleva, Eliyas and Petrov, 2000) studied the power spectra of 80 time series recorded at wide range of control parameters at the oxidation of HCOOH over supported Pd catalyst. It is found out that they exhibit a persistent behavior, namely: all of them comprise a continuous band that uniformly fits the shape $1/f^{\alpha(f)}$ where $\alpha(f) \to 1$ at $f \to 1/T$, $T$ is the length of the time series, and $\alpha(f)$ monotonically increases with the frequency. In our previous paper (Koleva and Covachev, 2001) it is proven that the presence in the power spectrum of a continuous band of the shape $1/f^{\alpha(f)}$ is a distinctive property of the bounded irregular sequences. The mechanism that drives bounded fluctuations is the diffusion-induced noise.

It is worth noting that the features of the power spectra obtained in (Koleva, Eliyas and Petrov, 2000) call for a new mechanism of fluctuations. On the one hand, the preparation of the system is such that the fluctuations are supposed negligible which ensures the kinetics evolution to be described by rate equations. However, the solution of neither system of ordinary or partial differential equations (rate equations) can comprise both a discrete and a continuous band. Then, if the discrete band comes from a limit cycle, what is the origin of the continuous one. On the other hand, the established so far sources of internal fluctuations gives rise to macroscopic effects only around the bifurcation points. However, there is no continuous band of the shape $1/f^{\alpha(f)}$ among these effects. Further, neither general model for explanation of the ubiquitous $1/f$ noise phenomenon is relevant to the surface reactions (McWorther, 1957). Moreover, the shape of the power spectra in the models such as the famous SOC, introduced by Bak et al (Bak, Tang and Wiesenfeld, 1987), strongly depends on the particularity of the incremental statistics and on the dimensionality of the system.

The coexistence of a discrete and continuous band of the above shape is an immediate result of eqs.(1). The deterministic part, eqs.(2), can give rise to a limit cycle at certain values of the control parameters while the stochastic part (which is a bounded irregular sequence) always produces a continuous band of the shape $1/f^{\alpha(f)}$ (Koleva and Covachev, 2001). The continuous band of the shape $1/f^{\alpha(f)}$ is evident in all 3 power spectra presented above. The deviations of $\alpha(f)$ from 1 become even visible for larger frequencies.

**Conclusions**

The goal of the present paper is the explicit evaluation of the features of the large-scaled fluctuations yielded by the diffusion-induced noise. The effective change of the control parameters caused by the any fluctuation can brings about sharp sudden changes of the system behavior in the course of the time even when the catalyst properties remain unchanged. They include: drastic change of the dynamical regime (i.e. a bifurcation), triggering of an adsorbate phase transition, change of the reaction route etc. The importance of the problem is enhanced by the inevitability of the diffusion-induced noise: it is persistent at each surface reaction at each value of the control parameters.

The major assumption put forth is that there is a diffeomorfism between the distance to a bifurcation point and the size of fluctuations. Therefore, the study of such features as size, duration and frequency for occurrence of the fluctuations are the first step in the study of the noise-induced bifurcations and other noise-induced phenomena. We carried out analytically the relations between the size, duration and the duration of the embedding intervals of any excursion. The dependence of the frequency for occurrence of a fluctuation of a given size $A$ on the particularities of the reaction mechanism is revealed through the power $\beta(A)$ in eq.(14). The proof that $\beta(A)$ always tends uniformly to 0.5 on increasing $A$ renders a gradual decrease of the influence of the incremental statistics (respectively the reaction mechanism) on the features of the large-scaled fluctuations (excursions).

It should be stressed that the boundedness introduces the "embedding" of each excursion in larger time interval. This does not allow the overlapping of the successive excursions and thus prevents the grow of the excursion amplitude to an arbitrary size. Thus, the "embedding" preserves the boundedness arbitrarily long-term.

The inevitability of the diffusion-induced noise and the fluctuations that it brings about gives rise to the question about a further development of a general frame of the possible effects that fluctuations can cause under more realistic constraints (e.g. self-organised patterns of adspecies, catalyst changes etc.). In particular, the construction of an appropriate feedback that can prevent further increases of a fluctuation above certain level is an important future task in providing long-term stable evolution of the industrial reactors.

Figure 1a

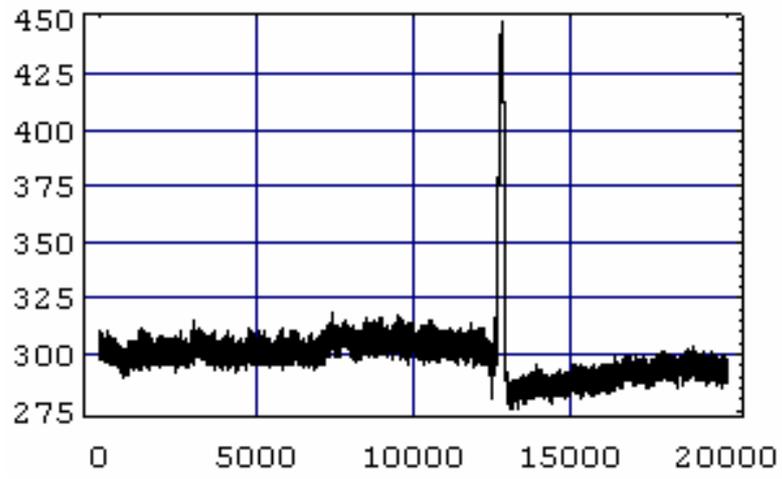

Figure 1b

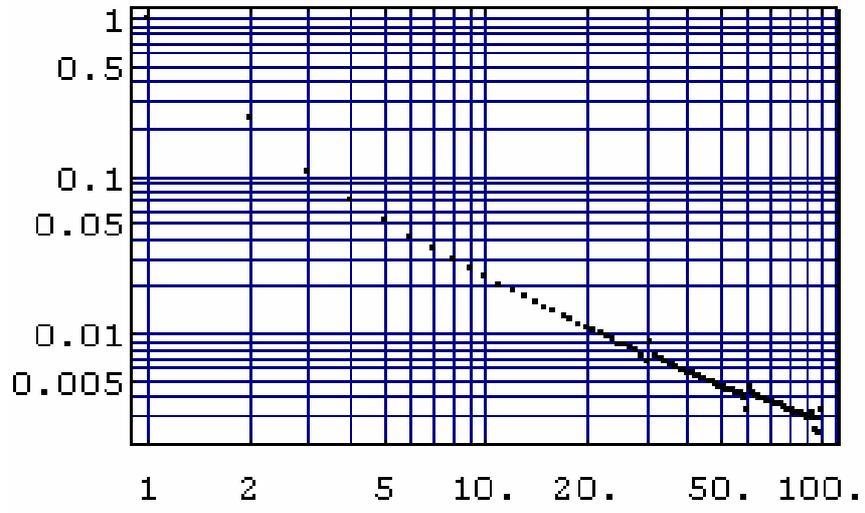

Figure 2a

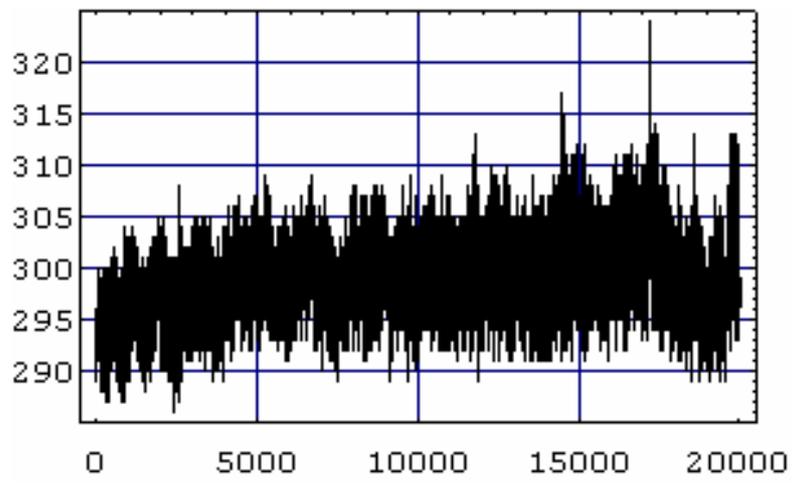

Figure 2b

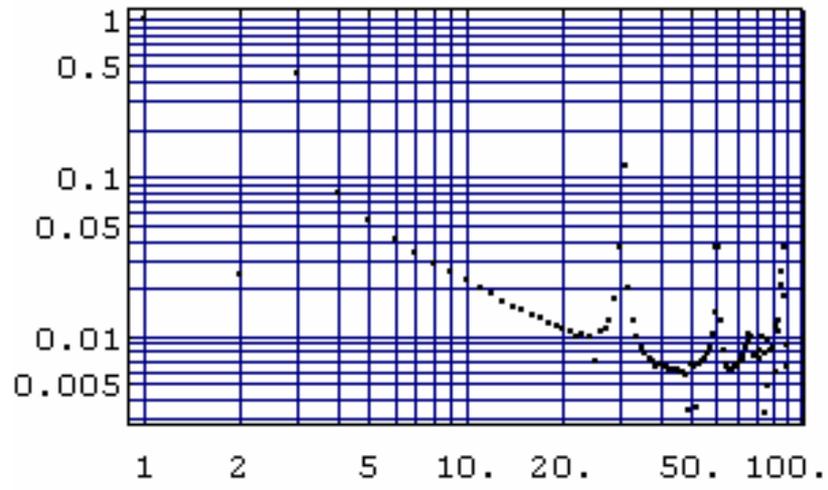

Figure 3a

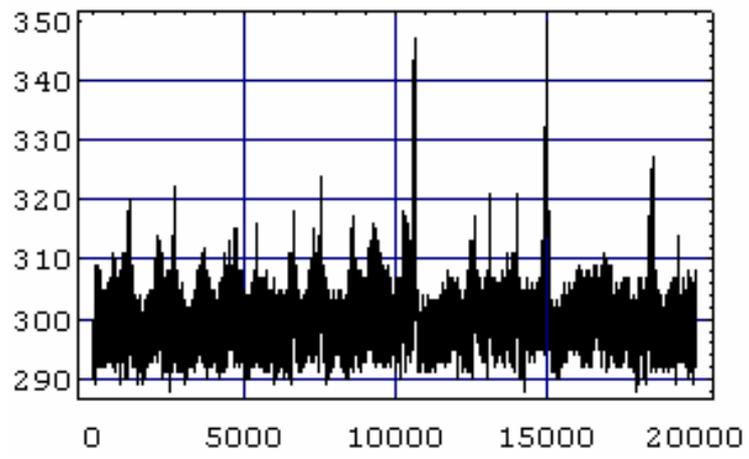

Figure 3b

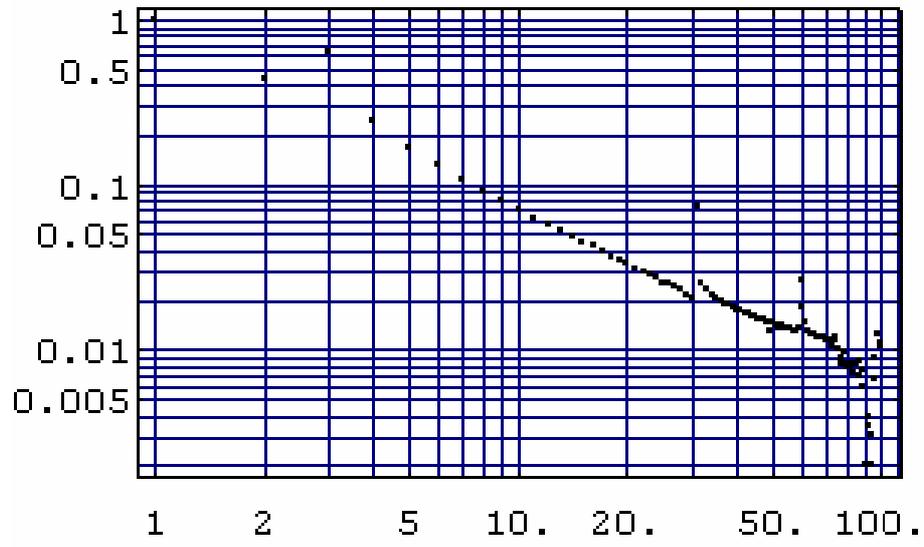

**Captures to the Figures**

Fig.1(a) A time series in relative units of the catalyst temperature variation (°K) in the course of time (sec) at the oxidation of HCOOH at a limit cycle: first part

Fig.1(b) A power spectrum in relative units of the catalyst temperature variations (°K) in the course of time (sec) at the oxidation of HCOOH at a limit cycle : first part

Fig.2(a) A time series in relative units of the catalyst temperature variation (°K) in the course of time (sec) at the oxidation of HCOOH at a limit cycle: second  part

Fig.2(b) A power spectrum in relative units of the catalyst temperature variations (°K) in the course of time (sec) at the oxidation of HCOOH at a limit cycle :  second part

Fig.3(a) A time series in relative units of the catalyst temperature variation (°K) in the course of time (sec) at the oxidation of HCOOH at a limit cycle: third  part

Fig.3(b) A power spectrum in relative units of the catalyst temperature variations (°K) in the course of time (sec) at the oxidation of HCOOH at a limit cycle :  third  part